# Ultrafast light-sheet optical tweezers for *in situ* parallelized biomechanical characterization of cells and soft tissues

**Krishangi Krishna,[a,b] Gannon Lemaster,[a] Jieyi Xu,[a] Zhaowei Jiang,[a,b] Carlonia G. Casas,[a,b] Zahra Ahmed,[a,b] Stephanie Roser,[a,b] Kareen L. K. Coulombe,[a,b] Anita Shukla,[a,b] Vikas Srivastava,[a,b] Kimani C. Toussaint, Jr.[a,b,c*]**

[a] Brown University, School of Engineering, Providence, RI 02912, USA

[b] Brown University, Institute for Biomedical Engineering and Medicine, Providence, RI 02912, USA

[c] Brown University, Center for Digital Health, Providence, RI 02903, USA

**Abstract**. Quantitative characterization of the mechanical properties of cells and tissues is essential for understanding disease progression and tissue regeneration. Optical tweezers (OT) enable the direct application of biologically relevant forces; however, OT has been limited to single axial indentations of cells, thereby restricting throughput. Furthermore, the use of quadrant photodiodes is insufficient for assessing the large displacements required for biomechanical characterization of tissues. We present light-sheet optical tweezers as a force transducer (LOFT), an approach that improves the throughput by at least 3× through simultaneous multiparticle trapping and parallelized characterization under sub-nN forces. LOFT is achieved by uniquely integrating light-sheet illumination for extended trapping, femtosecond-pulsed lasers to augment the optical gradient force, and videography-based particle tracking for observation of force transduction. The platform is validated through single-cell indentation experiments. We then apply LOFT to myocardial tissue, revealing significant biomechanical differences between healthy and infarcted regions; the interpretation of which is further supported by quantitative multiphoton imaging using the same optical source and platform. This work represents the first demonstration of OT for the mechanical testing of intact soft tissues, and establishes LOFT as a versatile, multifunctional platform for high-throughput, minimally invasive, mechanical characterization of complex biological systems *in situ*.

***Kimani C. Toussaint, Jr.**, E-mail: kimani_toussaint@brown.edu

## 1 Introduction

The ability to quantitatively characterize the mechanical properties of biological cells and tissues is essential for understanding migration, proliferation, differentiation, extracellular matrix (ECM) remodeling, and disease progression.[1-8] Indeed, establishing direct relationships between tissue mechanics and underlying microstructural organization remains an important challenge in mechanobiology. Mechanical remodeling is frequently accompanied by substantial changes in ECM organization, collagen alignment, and cellular metabolic activity. For example, pathological remodeling following myocardial infarction is associated with fibrosis, collagen deposition, and disruption of the highly anisotropic myocardial fiber network, ultimately leading to altered tissue stiffness and mechanical behavior.[9,10] Alterations in elastic and viscoelastic behavior have emerged as important biomechanical biomarkers associated with pathological states including cancer, fibrosis, and cardiovascular disease.[11-15] Thus, there exists an increasing demand for experimental platforms capable of high-throughput mechanical characterization while resolving structural organization for understanding structure-function relationships in complex biological systems. Such platforms should offer high spatial precision under biologically relevant loading conditions while preserving cellular viability and native tissue architecture.

Numerous techniques have emerged over the past decades to probe the mechanical properties of biological materials spanning molecular systems, single cells, and multicellular tissue assemblies. Existing methodologies vary considerably in their operating principles, applied force regimes, temporal resolution, throughput, and suitability for dynamic *in situ* measurements. Bulk mechanical testing (uniaxial tensile and compression) provides macroscopic material properties using coarse forces ranging from 0.2 – 2000 kN, but lacks spatial resolution (~mm), and is inherently destructive with poor repeatability.[16-18] Inference-based approaches (e.g., traction force microscopy, laser ablation, passive microrheology) often require extensive sample preparation and complex computational reconstruction.[19-22] Direct indentation methods (e.g., atomic force microscopy, instrumented nanoindentation) apply forces (100 pN – 100 nN) potentially damaging delicate samples. Furthermore, these platforms are generally low throughput, require additional integration of imaging modalities for sample visualization, and often suffer from low signal-to-noise ratio (SNR) at low loading conditions.[13-25] More fundamentally, many direct probing techniques cannot accurately recapitulate the mechanical environment experienced by cells *in vivo*, where mechanotransduction occurs under forces of only a few pN. Operation at substantially higher force regimes can induce irreversible deformation, cytoskeletal remodeling, or mechanotransductive responses that may alter the intrinsic mechanical properties being measured.[23] Furthermore, limited compatibility with high-speed imaging restricts real-time observation of dynamic biomechanical processes in native tissue environments. Optical coherence elastography (OCE) and Brillion microscopy provide label-free and non-destructive techniques for mechanical characterization of tissues. However, OCE performance can be limited by speckle noise, phase washout, and sensitivity to motion artifacts, while Brillion scattering remains limited to *ex vivo* and relatively static biological processes.[26-29]

Among minimally invasive biomechanical approaches, OT has emerged as particularly attractive due to its ability to apply and measure forces ranging from 0.01 - 100 pN while simultaneously tracking particle displacements with nanometer spatial precision. These force regimes closely match those encountered by cells *in vivo*, making OT uniquely suited for probing molecular motors, cytoskeletal mechanics, receptor-ligand interactions, intracellular transport, and membrane fluctuations under biologically relevant loading conditions.[30-34] An additional advantage of OT is its compatibility with a wide range of optical imaging modalities, enabling simultaneous mechanical characterization and structural visualization within a single microscope platform. Despite these advantages, conventional OT systems remain fundamentally constrained to single-particle trapping with a Gaussian beam, localized force application, and the use of quadrant photodiodes (QPD) for position detection. Of note is that QPD-based force measurements are highly sensitive to optical alignment and beam aberrations, and are generally optimized for single-particle tracking over small displacement ranges approximating 200 nm.[35] As a result, biomechanical measurements must typically be performed sequentially, limiting throughput and restricting the ability to rapidly characterize large or mechanically heterogeneous specimens. These limitations become increasingly significant when interrogating mechanically heterogeneous tissues, where local mechanical properties may vary substantially across micron-scale regions.[36] Moreover, quantitative tissue biomechanics often necessitates indentation depths extending from hundreds of nm to several μm, exceeding the displacement range for which conventional QPD-based position detection is typically optimized. Another concern is that most OT-based force transduction platforms employ axial indentation geometries, wherein the trapped probe is displaced along the optical axis (parallel to the direction of the illumination beam). However, axial interrogation comes with two primary challenges. The first is that it is well-known that cells use focal adhesion to anchor to substrates. Thus, mechanical assessment obtained by axial loading with a force probe cannot deconvolve the substrate mechanical contribution from that of the cell without a priori information.[37] The second is that the approach is particularly sensitive to axial drift or defocusing because the OT system and QPD detection share the same optical path. Furthermore, optical imaging of indented materials is often carried out in a channel that is physically decoupled from that of the OT force probe, and thus co-registration of inferred biomechanical information with visualized data is challenging.[38] Consequently, the accuracy of current OT platforms for biomechanical characterization is significantly constrained by the topology of the experimental setup. Thus, there is a need for new biomechanical instrumentation with improved accuracy and throughput compared to the state-of-the-art.

There are three potential approaches to increasing the throughput of OT-based biomechanical testing platforms. The first is to time-multiplex a single-focused beam. However, the rate-limiting step is the time required to indent the specimen, which will usually be on the order of 15 sec, thereby resulting in a sequential rather than parallelized process.[39] The second is to create a discrete array of focused points, either 1D or 2D, using either microlens arrays or holographic projection. This approach is much more favorable than the first, but is ultimately hampered by power, which is divided by the number of elements in the array.[40] An intriguing, third method uses light-sheet (LS) illumination, where a single focus is spatially extended along one of the transverse axes, thereby, in principle, permitting multiple locations of a specimen to be interrogated simultaneously. Light-sheet optical tweezers has enabled 2D cell trapping and sample rotation while simultaneously performing high-resolution fluorescence imaging, typically using two separate continuous-wave (CW) lasers for OT and fluorescence excitation.[41-43] A major challenge, however, is that the optical gradient force along the minimally focused dimension (short axis) is severely reduced compared to a standard optical trap generated by a circularly symmetric, focused Gaussian beam; such an approach is therefore buttressed by using high CW-laser input power (> 50 mW) and by pushing particles against a rigid surface, thereby limiting its practical viability as a biomechanical testing platform.[41,42] Nevertheless, the use of LS-illumination provides an interesting strategy towards achieving our goal of increasing the throughput—one where the limitations can be addressed by judiciously integrating the use of a femtosecond-pulsed optical source in combination with videography-based particle tracking.

We previously demonstrated that femtosecond laser-assisted selective holding with ultra-low power (FLASH-UP) enables enhanced particle confinement at substantially reduced average powers (sub–1 mW) through the generation of an auxiliary force that acts synergistically with the optical gradient force induced by ultrafast laser pulses. Specifically, the trap stiffness afforded by FLASH-UP is 5× higher than CW-OT [44]. This low-average-power regime is particularly advantageous for biological applications because it minimizes photothermal damage and perturbation of native cellular processes while still enabling robust optical force generation.[44,45] Thus, FLASH-UP could be used to enhance the trap stiffness for light-sheet-based OT. We previously demonstrated proof-of-concept experiments of stable, 3D trapping of multiple 2-μm diameter silica particles with FLASH-UP and light-sheet illumination at an average power as low as 1 mW.[46] In addition, increased throughput by parallelization can be achieved through the use of videography for particle tracking as long as the appropriate image-processing algorithms can accurately identify the boundaries between the trapped objects and the biological specimen.

In this work, we present light-sheet optical tweezers as a force transducer (LOFT), an approach that increases the throughput of OT-based mechanical testing by at least threefold. LOFT is achieved by uniquely integrating three distinct features: light-sheet illumination for an extended and uniform trapping region, high-repetition-rate, femtosecond-pulsed lasers to augment the optical gradient force, and videography-based particle tracking to parallelize the observation of force transduction from trapped beads to the biological material. The platform is first validated through single-cell indentation experiments on cells with differing physiological functions and pathogenic states. The system is subsequently applied to healthy and infarcted myocardial tissues, where indentation measurements are directly correlated to quantitative, multiphoton microstructural information. To our knowledge, this is the first demonstration of OT-based non-destructive, *in situ*, simultaneous indentation and visualization performed directly on tissue specimens within biologically relevant force regimes. Finally, LOFT enables multipoint indentation to simultaneously interrogate spatially heterogeneous tissue regions within a single experiment, demonstrating parallelized biomechanical characterization under non-destructive force-loading conditions.

## 2 Materials and Methods

### *2.1 Experimental Setup*

Figure 1(a) shows the experimental setup used. An ultrafast laser source (Insight X3, Spectra-Physics) operating at a central wavelength of 800 nm, with a pulse duration of 120 fs and repetition rate of 80 MHz, is employed. Near-infrared light is used to minimize photodamage to biological samples as water and other biological components exhibit relatively low optical absorption in this spectral region.[47,48] The laser is first expanded and spatially filtered to ensure a $TEM_{00}$ profile prior to entering the microscope system. Precise control of the average power at the sample plane is achieved using a half-wave plate mounted on a motorized rotational stage (Thorlabs) in conjunction with a linear polarizer (LP). A quarter-wave plate is subsequently introduced to generate circularly polarized light. The beam is then directed onto a 2D galvanometer scanning system (Thorlabs) before being coupled into an inverted Olympus IX83 microscope (Evident) through a dichroic mirror (DM) and focused by a 100×/1.3 numerical aperture (NA) oil-immersion objective lens. To generate a LS, a cylindrical lens (CL) is incorporated into the optical path. The CL is accessed through a flip mirror (FM) positioned immediately before the entrance aperture of the microscope objective, and mounted on a translational stage to facilitate precise beam alignment. This arrangement enables rapid switching between Gaussian and LS beams without substantial modification of the optical setup. For real-time visualization during indentation experiments, brightfield (BF) illumination from the microscope lamp is used. Backscattered light from the sample plane is filtered using a short-pass filter to improve image contrast and suppress background signal during image acquisition. Time-lapse image sequences of the indentation process are recorded using an sCMOS camera.

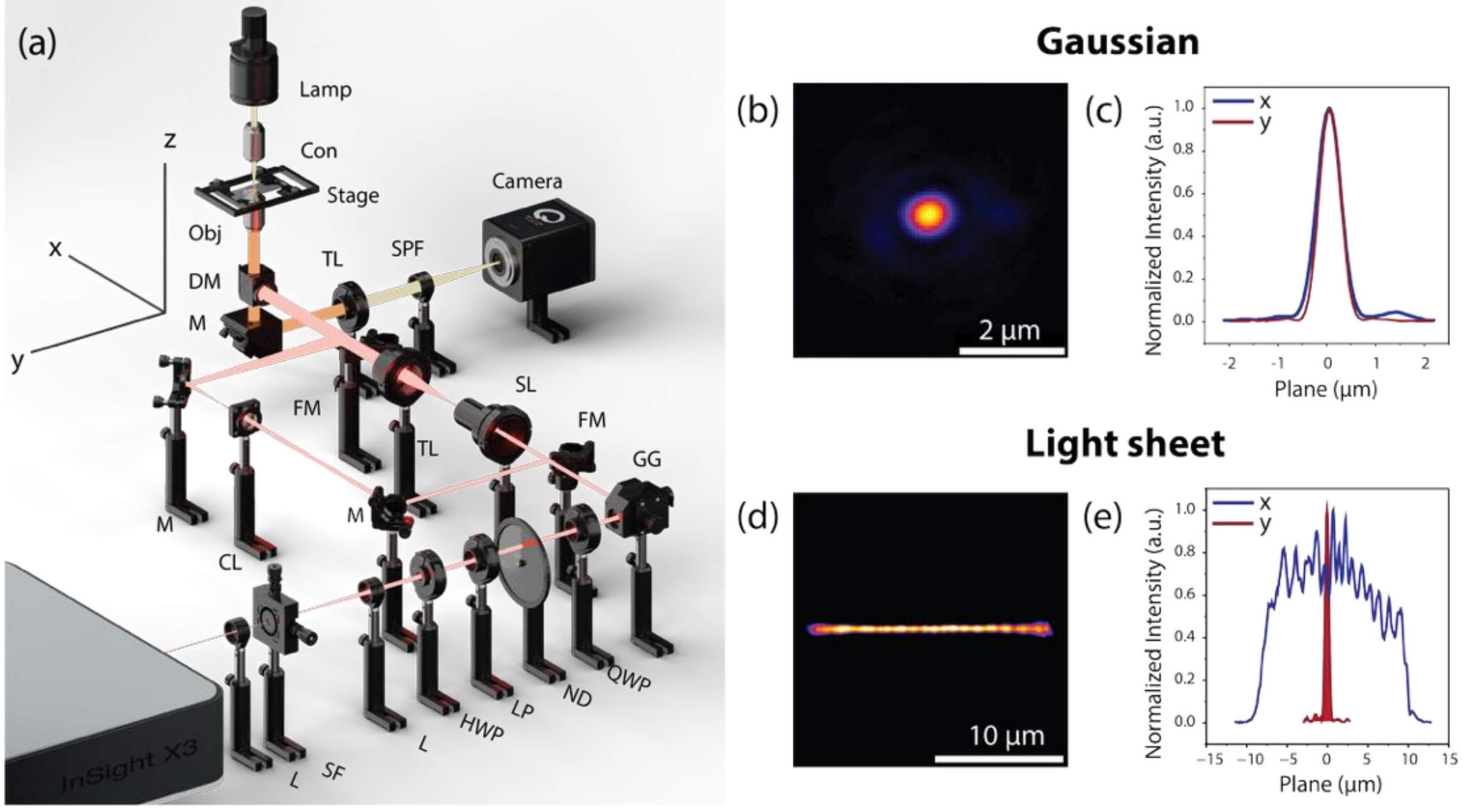

**Fig. 1.** Experimental arrangement. (a) Schematic illustration of LOFT. L; lens, SF; spatial filter, HWP; half wave plate, LP; linear polarizer, ND; neutral density, QWP; quarter wave plate, M; mirror, FM; flip mirror, SL; scan lens, TL; tube lens, CL; cylindrical lens, DM; dichroic mirror, SPF; short pass filter. (b) Experimentally obtained beam profile of Gaussian. (c) Normalized transverse intensity profile along *x* (blue) and *y* (red). (d) Transverse beam profile of LS. (e) Corresponding normalized intensity profiles along *x* (blue) and *y* (red).

The experimentally obtained Gaussian beam profile and its corresponding normalized intensity cross-sections along the *x*- (blue) and *y*-axes (red), demonstrating a focal spot size of ~ 400 nm, are presented in Figs. 1(b,c), respectively. In contrast, the LS beam profile and its corresponding normalized transverse intensity distributions along the *x*- (blue) and *y*-axes (red), are shown in Figs. 1(d,e), respectively, with dimensions of approximately 20 µm × 1 µm, along *x*- (blue) and - and *y*-axes (red) respectively.

*2.2 Sample Preparation*

Murine macrophages (RAW 264.7) and murine fibroblasts (NIH 3T3) are obtained from American Type Culture Collection (ATCC), Manassas, VA. Both cell lines are seeded and passaged in Dulbecco's Modified Eagle Medium (DMEM) supplemented with 10% (v/v) fetal bovine serum (FBS) and 1% (v/v) penicillin-streptomycin (Cytiva, Marlborough, MA) and incubated at 37°C with 5% $CO_2$. Cells are seeded at a density of 50,000 cells/mL in 6-well tissue culture treated plates (Celltreat, Pepperell, MA), with each well containing a micro coverglass (12 mm in diameter, Electron Microscopy Sciences, Hatfield, PA) and DMEM supplemented with FBS. A 1:10 dilution series is performed across all wells. The cells are incubated at 37°C with 5% $CO_2$ overnight to allow attachment.

MCF-10A human breast epithelial cells (ATCC) are cultured in complete DMEM/F12 (Gibco) supplemented with 15 mM HEPES, 5% (v/v) horse serum (Gibco), 1% (v/v) penicillin-streptomycin (Gibco), 20 ng/mL EGF (Peprotech), 0.5 mg/mL hydrocortisone (MilliporeSigma), 10 μg/mL insulin (MilliporeSigma), and 100 ng/mL cholera toxin (MilliporeSigma). MCF-7 human breast cancer cells (ATCC) are maintained in DMEM/F12 (Gibco) containing 15 mM HEPES, 10% (v/v) FBS (Gibco), and 1% (v/v) penicillin-streptomycin (Gibco). All cells are expanded in tissue culture-treated T-25 (Fisher Scientific) at 37°C in a humidified incubator with 5% CO2. Media is replaced every 48 hours, and cells are passaged when they reach 80% confluence. Experiments are conducted using cells between passages 4 and 6. German glass coverslips (#1.5, 20 mm; Electron Microscopy Sciences) are acid-cleaned in 1 M hydrochloric acid (MilliporeSigma) for 4 h, followed by four washes in sterile molecular-grade deionized (DI) water (Fisher Scientific). Coverslips are then dipped in 100% ethanol, transferred to 6-well plates (Corning), and UV-sterilized for 30 min. Following three additional washes in sterile DI water, coverslips are coated with 10 μg/cm2 fibronectin (MilliporeSigma) for 1 hour and rinsed once with phosphate buffer saline (PBS). Cells are seeded onto the coated coverslips at a density of 15,000 cells/cm2 in 600 μL of suspension. After overnight incubation, 1.4 mL of additional media is added to each well prior to subsequent experiments. Wells containing single cells attached to the coverglass are then removed from the medium, and supplemented with a 25 μL aliquot of 2-μm diameter dielectric microspheres, and covered with a gasket (Thermofisher).

All animal procedures are approved under IACUC #23-02-0003 at Brown University. A rat model of ischemia/reperfusion myocardial infarction was produced as described previously.[49,50] Briefly, 8-week-old male wild type Sprague-Dawley rats are anesthetized with 2-3% isoflurane followed by 100 mg/kg ketamine and 10 mg/kg xylazine then intubated and placed on a ventilator. A longitudinal incision is used to perform a thoracotomy and expose the heart. A 7-0 polypropylene suture is tied over tubing to occlude the left anterior descending coronary artery for 60 min using visual assessment of the blanched area at risk to confirm positioning. After 60 min, the suture is removed to enable reperfusion and the heart is harvested, fixed in 4% paraformaldehyde overnight, then stored short-term in PBS before processing. Hearts are sliced into 2 mm transverse sections, submerged in 30% sucrose for 72 hrs or until sections have sunk, then embedded and frozen in optimal cutting temperature compound. Frozen tissues are sectioned at 10-μm thickness onto coverslips, drop-cast with a 40-μL suspension of 10-μm diameter dielectric microspheres and 10% BSA, and enclosed with a gasket.

*2.3 Videography-based particle tracking algorithm*

One-minute, time-lapse videos of the trapped particles within a field of view (FoV) of 9.36 x 9.36 $\mu m^2$ and a frame rate of 1000 frames per second are recorded. OpenCV[51] is used to convert images from RGB to grayscale. To reduce high-frequency noise, a Gaussian blur is applied by convolving each image with a 5x5 kernel ($\sigma = 1$). Particle positions are then extracted using a circular Hough transform (HT). Detection parameters, including the high-hysteresis threshold of the edge detector, the accumulator threshold for circle centers in parameter space, and the minimum and maximum detectable radii, are tuned to ensure robust particle identification over the full time-lapse. The time-dependent $x$ and $y$ coordinates of the trapped particle extracted via the circular HT are used to construct position histograms with a predefined bin width. The cutoff frequency is extracted by fitting the experimentally measured data to the power spectrum using a nonlinear least squares algorithm up to frequency $f_{hf}$ (> 100 Hz) in order to exclude high-frequency noise contributions. Additional details regarding trap stiffness ($\kappa$) extraction are provided in Refs 44, 45.

HT-image processing is also employed to extract the time-dependent transverse coordinates of the probe and cell during indentations, with contrast enhancement performed using contrast-limited adaptive histogram equalization instead of Gaussian smoothing for indentation experiments. To reduce incorrect spatial detection of the biological sample and trapped particle, an additional criterion is applied when multiple candidate circles are detected. For the bead, which produces more false positive circle detection due to its smaller radius, the candidate closest to its initial position is chosen. For biological samples, the candidate closest to the position of the previous time step is chosen. Tissue indentation experiments require a slightly adapted measurement strategy due to the irregular geometry of the tissue surface. The tissue boundary in contact with the bead is manually identified and used as a reference for the predetermined displacement.

### *2.4 Hertz Model*

The Hertz contact model is one of the most widely utilized frameworks for extrapolating biomechanical properties from AFM- and OT-based indentation experiments.[23,38] The model describes the interaction between a rigid spherical indenter and an isotropic, homogeneous, semi-infinite elastic material undergoing small deformations. Under these assumptions, the force–indentation relationship can be used to estimate the apparent elastic modulus ($E$), which characterizes the material's resistance to deformation under an applied load [52, 53], and is given by

$$E = \frac{3(1-\nu^2)}{4R^{1/2}\delta^{3/2}}F_{max}, \tag{1}$$

where $F_{max}$ represents the maximum force applied leading to an indentation depth of $\delta$ on a sample with Poisson's ratio $\nu$. Biological tissues and cells are commonly reported to possess Poisson ratio values ranging from 0.4 to 0.5 [38, 54]. Additional assumptions of the Hertz model include frictionless contact, negligible adhesive interactions between the probe and sample surface, linear elastic material behavior, and the absence of significant sample stratification or viscoelastic relaxation during indentation [38]. It also assumes that the indenter is significantly stiffer than the sample, such that deformation of the probe itself is negligible relative to deformation of the biological specimen. Importantly, this model is generally considered valid when the $\delta$ remains less than approximately 10% of the sample thickness along the indentation axis, thereby minimizing substrate effects and large-strain nonlinearities.[53,54]

Although biological systems do not fully satisfy all Hertzian assumptions due to their intrinsic heterogeneity, viscoelasticity, and structural complexity, the model remains widely employed as an effective first-order approximation for comparative biomechanical analysis.[38] In the context of this work, the primary objective is to establish relative mechanical differences between distinct cell populations and tissue states under consistent experimental conditions. Thus, the application of the Hertz model provides a justified and standardized framework for comparative evaluation of $E$ across the investigated biological systems.

*2.5 Statistical Analysis*

Experiments are conducted on 3 separate samples (biological replicates) and repeated 3 times (technical replicates) for reproducibility and statistical consistency. Statistical analyses are conducted on OriginPro. For cell experiments, since two separate populations are considered, an unpaired two-tailed Student's t-test is performed with a 95% confidence interval. For tissue indentation experiments, a paired two-tailed t-test is evaluated to determine statistical significance as experiments are conducted within the same sample.

*2.6 2D Fourier Analysis*

Each image is segmented into smaller interrogation windows of 5.2 x 5.2 $\mu m^2$ and the corresponding 2D Fourier spectrum is computed for each region. Regions exhibiting a well-defined dominant orientation within the frequency domain are classified as anisotropic; whereas regions lacking a predominant directional component are classified as isotropic. Areas with negligible signal intensity are categorized as dark regions, as insufficient structural information is present to reliably determine fiber orientation. Additional implementation details and algorithmic specifications of the 2D-fast Fourier transform (FFT) analysis framework are provided in Refs. 55-58.

## 3 Results

*3.1 Probing Cellular Mechanics Across Biological Function and Disease States*

An average power of 1 mW at the sample plane is utilized for single-cell indentation experiments with the Gaussian beam. Initially, a dielectric probe (green) is trapped and maneuvered into close proximity to the target cell (red) [Fig. 2(a)]. Once the desired indentation plane is identified, the microscope stage is translated at a velocity of 1 μm/s with an acceleration of 1 $\mu m/s^2$ over a total displacement of 3 μm, followed by retraction using identical motion parameters. The stage displacement (SD; red) is determined by tracking the displacement of the cell within the FoV, while the probe/ bead displacement (BD; green) is extracted using the particle-tracking algorithm described above. The indentation depth (blue) is determined as $\delta = SD - BD$. The optical force exerted on the sample (orange) is calculated using Hooke's law ($F = \kappa \Delta x$) [Fig. 2(b)] [23].

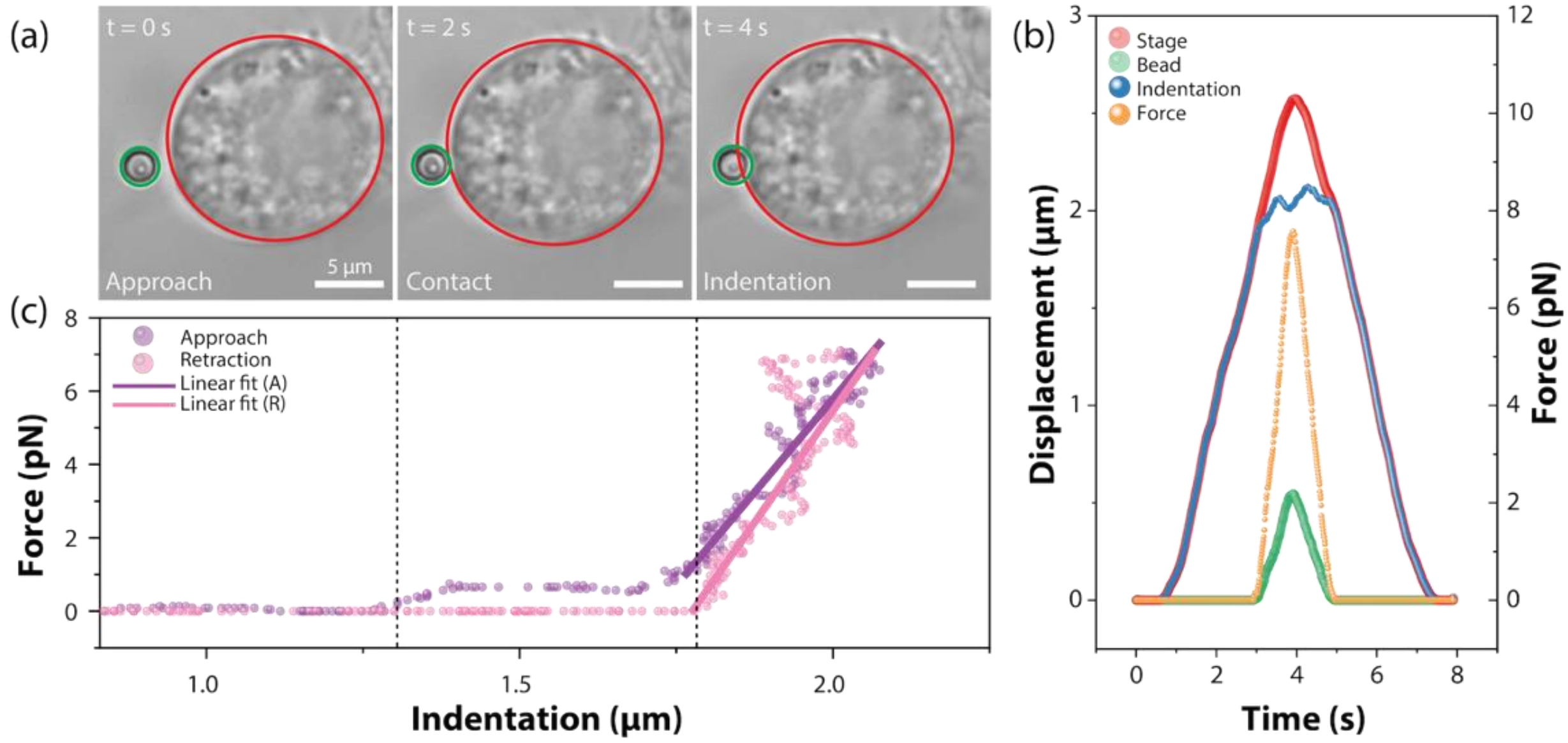

**Fig. 2.** Biomechanical characterization pipeline. (a) Representative BF time lapse demonstrating optical indentation of a macrophage cell (red) using a trapped dielectric microsphere (green). (b) Corresponding stage displacement (red), bead displacement (green), calculated indentation depth (blue), and applied force (orange). (c) Force-indentation curve illustrating approach (purple) and retraction (pink).

The corresponding force–indentation curve is shown in Fig. 2(c), from which three distinct experimental phases can be identified: approach, contact, and indentation. During the initial approach phase (purple), a small increase in force is observed prior to contact with the sample. This behavior is likely attributable to hydrodynamic drag and viscous resistance arising from movement of the probe through the surrounding fluid medium.[59] Upon contact, the force increases approximately linearly with indentation depth for both approach and retraction (pink). $E$ is determined from the linear regions of the force–indentation response.

The indentation platform is first evaluated for its ability to distinguish between cell types possessing inherently different biomechanical functions. Macrophages and fibroblasts are selected as representative model cells due to their well-established differences in physiological role and expected mechanical behavior. The average $\kappa$ of 2-μm diameter silica particles are measured to be 8.43 pN/ μm and 8.35 pN/ μm for macrophage and fibroblast experiments, respectively. The average maximum forces applied to the cells are 7.17 ± 0.34 pN for macrophages and 7.18 ± 0.31 pN for fibroblasts. Albeit the nearly identical applied force magnitudes, macrophages exhibit an average $\delta$ of 536 ± 227.32 nm (Video 1), whereas fibroblasts reveal a significantly smaller deformation of 215 ± 29.14 nm (Video 2) [Fig. 3(a)]. Macrophages display an $E$ of 12.53 ± 6.56 Pa, while fibroblasts show a significantly higher modulus of 41.87 ± 7.44 Pa [Fig. 3(b)]. Statistical tests reveal a $p$-value of 0.0069 supporting the hypothesis that the proposed platform can sensitively differentiate cell populations with distinct structural and functional phenotypes.

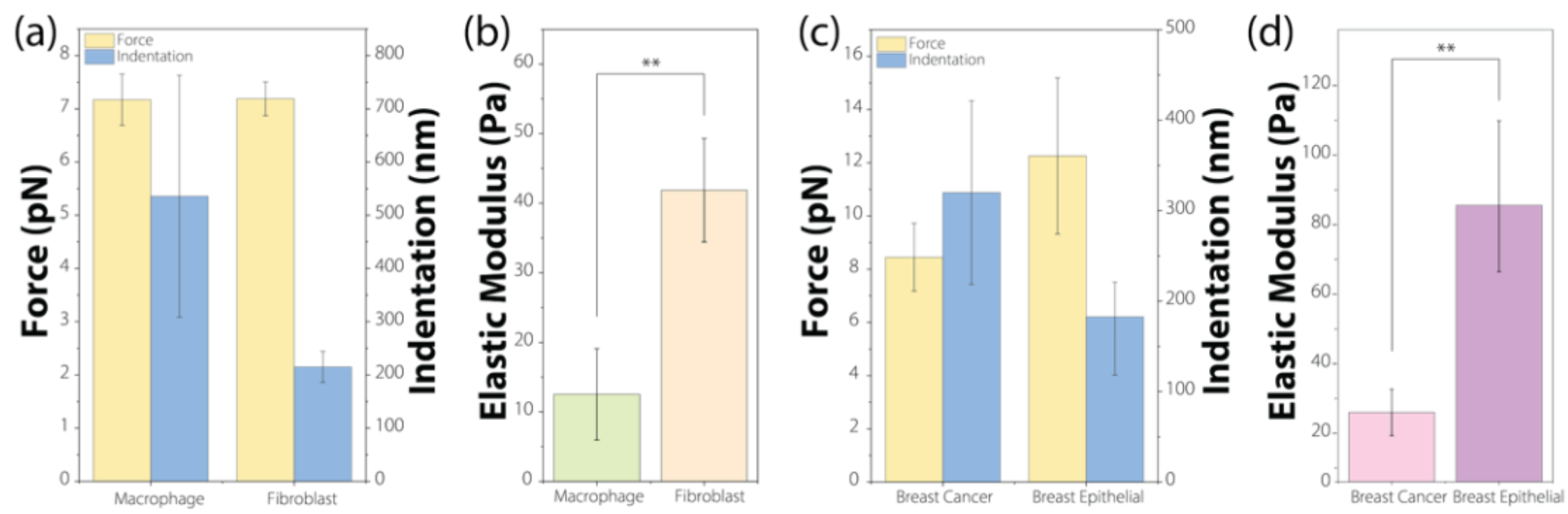

**Fig. 3.** Single indentation results of cells. (a) Comparison of $\delta$ under comparable force applied for macrophages and fibroblasts. (b) Extrapolated $E$ for macrophages and fibroblasts. (c) Comparison of $\delta$ measured for healthy breast epithelial and cancer cells. (d) Calculated E for breast epithelial and cancer cells. Note ** represents significant differences indicated by a Student's t-test ($n$ = 3) by $p < 0.01$.

The system is next applied to investigate cells in different pathological states by comparing healthy breast epithelial cells with breast cancer cells. The average $\kappa$ values of 2-μm diameter polystyrene (PS) particles are measured to be 13.87 pN/ μm for breast cancer cells and 11.89 pN/ μm for breast epithelial cells. PS is chosen over silica microspheres to minimize bead-cell adhesion. The average maximum force applied during indentation is 8.25 ± 1.27 pN for cancer cells and 12.26 ± 2.93 pN for epithelial cells. Although the epithelial cells undergo a larger applied force, breast cancer cells experience average indentations of 320 ± 101.39 nm (Video 3), whereas breast epithelial cells exhibit significantly smaller deformations of 181.48 ± 51.27 nm (Video 4) [Fig. 3(c)]. Breast cancer cells possess an $E$ of 28.44 ± 9.50 Pa, while breast epithelial cells reflect a substantially larger $E$ of 83.96 ± 1.49 Pa [Fig. 3(d)]. Statistical analysis results in a $p$-value of 0.0084, thereby demonstrating the sensitivity of the system in distinguishing cellular mechanical phenotypes associated with disease progression and pathological state.

### *3.2 Correlating Tissue Mechanics with Microstructural Changes*

For tissue indentation experiments with a single probe, an average power of 5 mW at the sample plane is used to displace the probe 7 μm. Average $\kappa$ is measured to be 63.31 ± 1.92 pN/ μm and 74.71 ± 1.70 pN/ μm for healthy and infarcted myocardium, respectively. The average forces applied during indentation are 105.63 ± 12.97 pN for healthy tissue and 139.57 ± 7.71 pN for infarcted tissue resulting in average $\delta$ of 379.89 ± 14.67 nm (Video 5) and 235.22 ± 31.4 nm (Video 6) for healthy and infarcted myocardium, respectively [Fig. 4(a)]. Despite experiencing larger applied forces, the infarcted tissues undergoes substantially smaller deformations, indicating a significantly stiffer mechanical response. Healthy myocardium yields an $E$ of 114.6 ± 8.75 Pa, whereas the infarcted myocardium displays a markedly higher modulus of 300.71 ± 66.75 Pa [Fig. 4(b)]. Statistical tests reveal a $p$-value of 0.032 illustrating that the platform is capable of sensitively differentiating between healthy and diseased tissue states through quantitative biomechanical characterization at sub-nN force regimes.

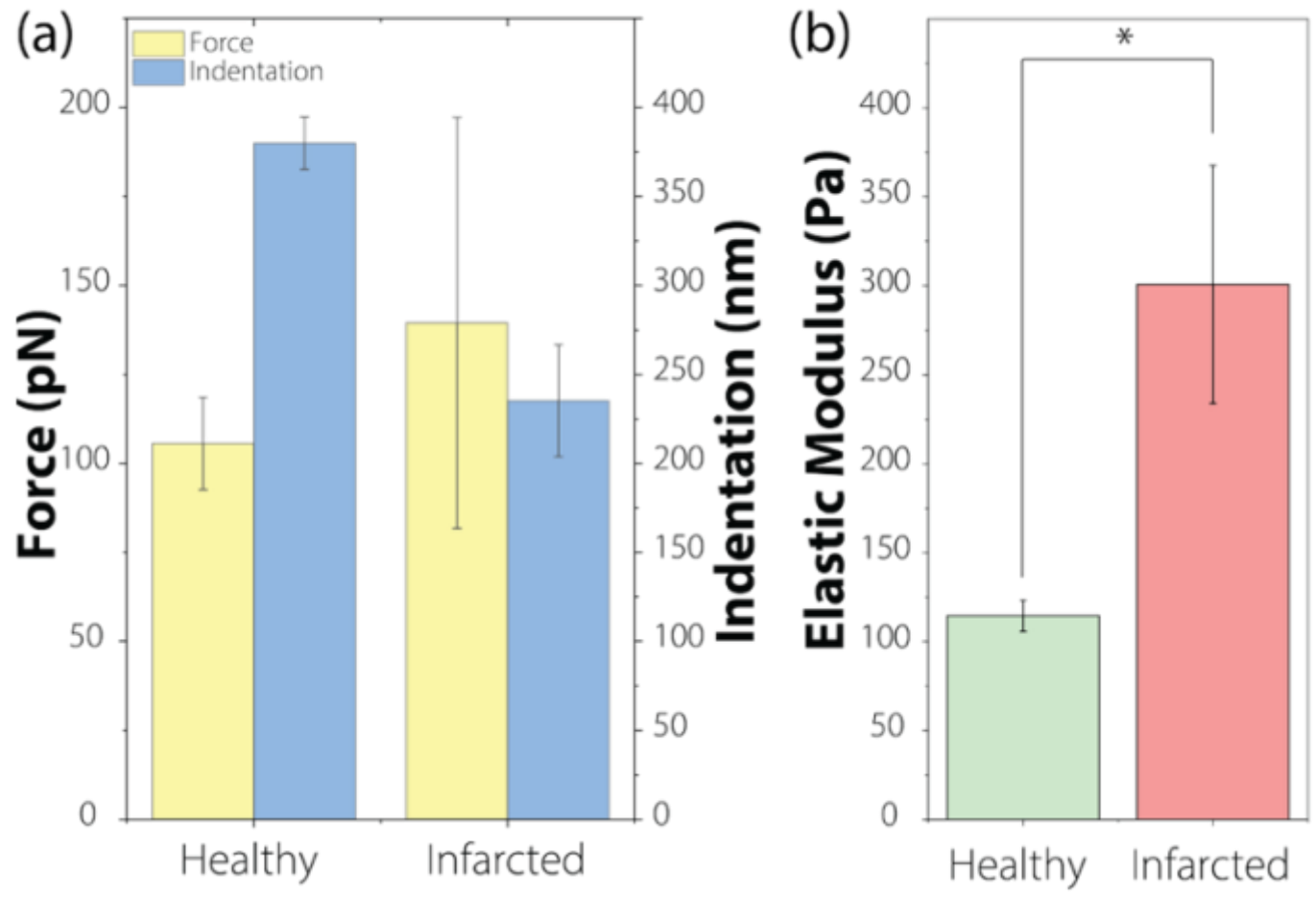

**Fig. 4.** Single indentation results of myocardium. (a) Comparison of $\delta$ measured for both tissue types using 2D Gaussian OT. (b) Estimated $E$ for healthy and infarcted tissue. Note * represents significant differences indicated by a Student's t-test (n = 3) by $p < 0.05$.

The healthy and infarcted myocardium is visualized through BF microscopy [Figs. 5(a,b)]. Two-photon excitation fluorescence (2PEF) images of the insets in Figs. 5(a,b) acquired through multiphoton excitation are shown in Figs. 5(c,d). A 2D spatial FFT analysis is performed on the acquired 2PEF images to evaluate the spatial-frequency content and directional organization of fibrous structures, enabling extraction of quantitative metrics including preferred fiber orientation and spatial-frequency distribution.

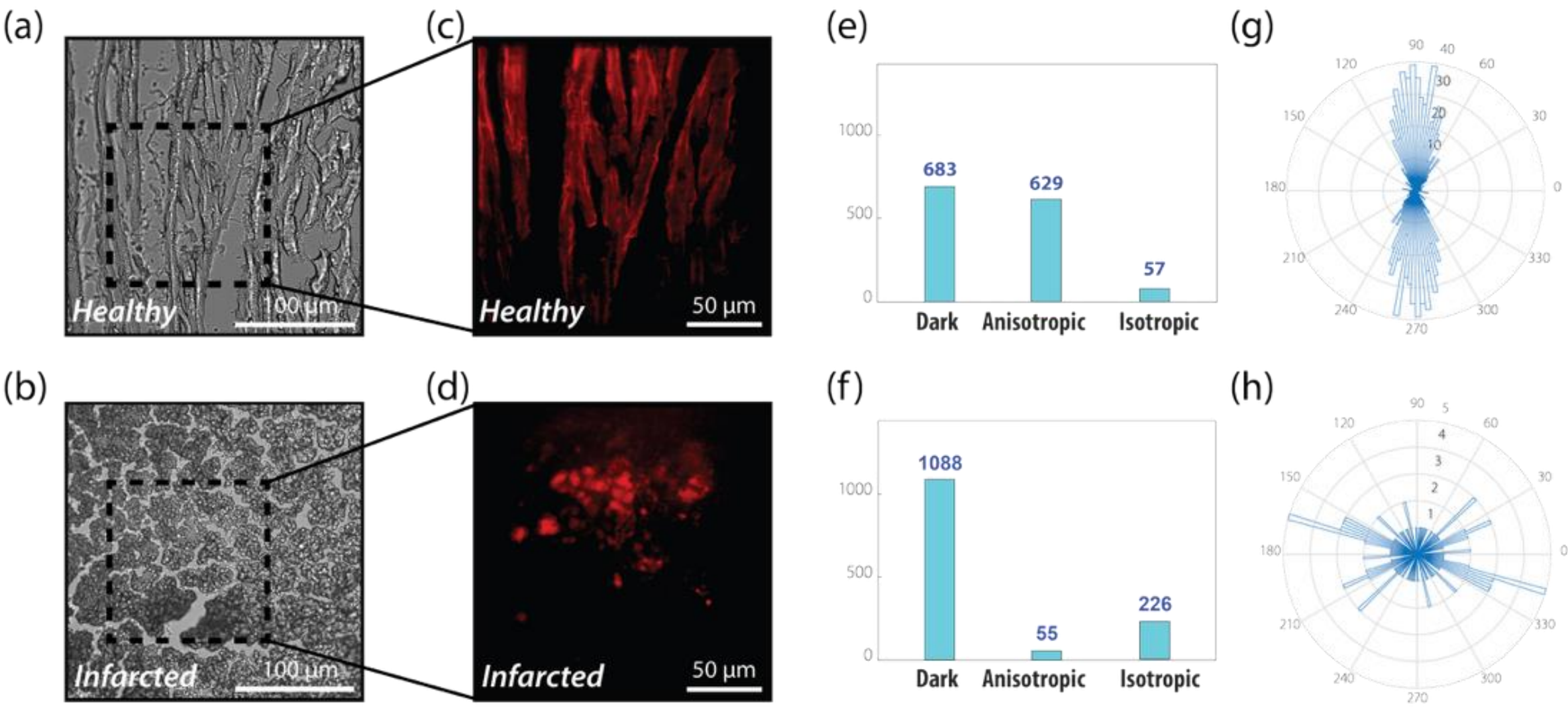

**Fig. 5.** Multiphoton imaging and analysis to determine tissue microarchitecture. (a,b) BF visualization of healthy and infarcted tissues. (c,d) 2PEF images from the insets of (a,b). (e,f) Spatial map illustrating anisotropic, isotropic and dark regions identified from analysis. (g,h) Polar histograms illustrating fiber orientation distribution and CV measurements.

The infarcted tissue displays nearly twice the number of dark regions compared to healthy tissue revealing substantial reductions in signal intensity and directional organization, consistent with the metabolic disruption, fibrosis, and collagen remodeling associated with post-infarction scar formation [60] [Figs. (e,f)]. In addition, the infarcted myocardium contains approximately 4× more isotropic regions relative to healthy tissue, indicating a substantial loss of organized fiber alignment following pathological remodeling. Healthy myocardial tissue reveals a preferred fiber orientation of 87.8° with a circular variance (CV) of 0.159, indicating a highly aligned and structurally coherent fiber network [Fig. 5(g)]. Conversely, infarcted tissue possesses an average orientation of 178.3° with a substantially larger CV of 0.499, reflecting reduced directional organization within the remodeled tissue matrix [Fig. 5(h)].

### 3.3 *Parallelized Biomechanical Characterization of Tissues*

LOFT experiments are conducted with an average power of 900 μW at the sample plane with three 10-μm diameter PS particles. We note that this power is at least 55× less than that reported in Refs 41-43, which we attribute to the additional leverage afforded by the FLASH-UP method. The LOFT indentation protocol consists of sequential particle loading, trap stabilization, controlled transport, and tissue indentation as described in the Supplementary Material, Figs S1 and S2, and shown in Video 7.

Three distinct experimental tissue conditions are investigated: homogenous regions consisting exclusively of (i) healthy cardiac tissue (Video 8), (ii) infarcted cardiac tissue (Video 9), and (iii) heterogeneous regions containing both healthy and infarcted tissue within the same FoV (Video 10). The trapped particles are labeled sequentially as P1–P3 from left to right along the LOFT axis. In a line-trap geometry, multiple particles simultaneously scatter and redistribute the incident optical field, resulting in non-uniform optical force distributions along the LS. Consequently, it cannot be assumed that all particles experience identical trapping conditions or equivalent restoring forces. The measured $\kappa$ along the $y$-axis, corresponding to the indentation direction, are presented for each experimental condition in Fig. 6(a).

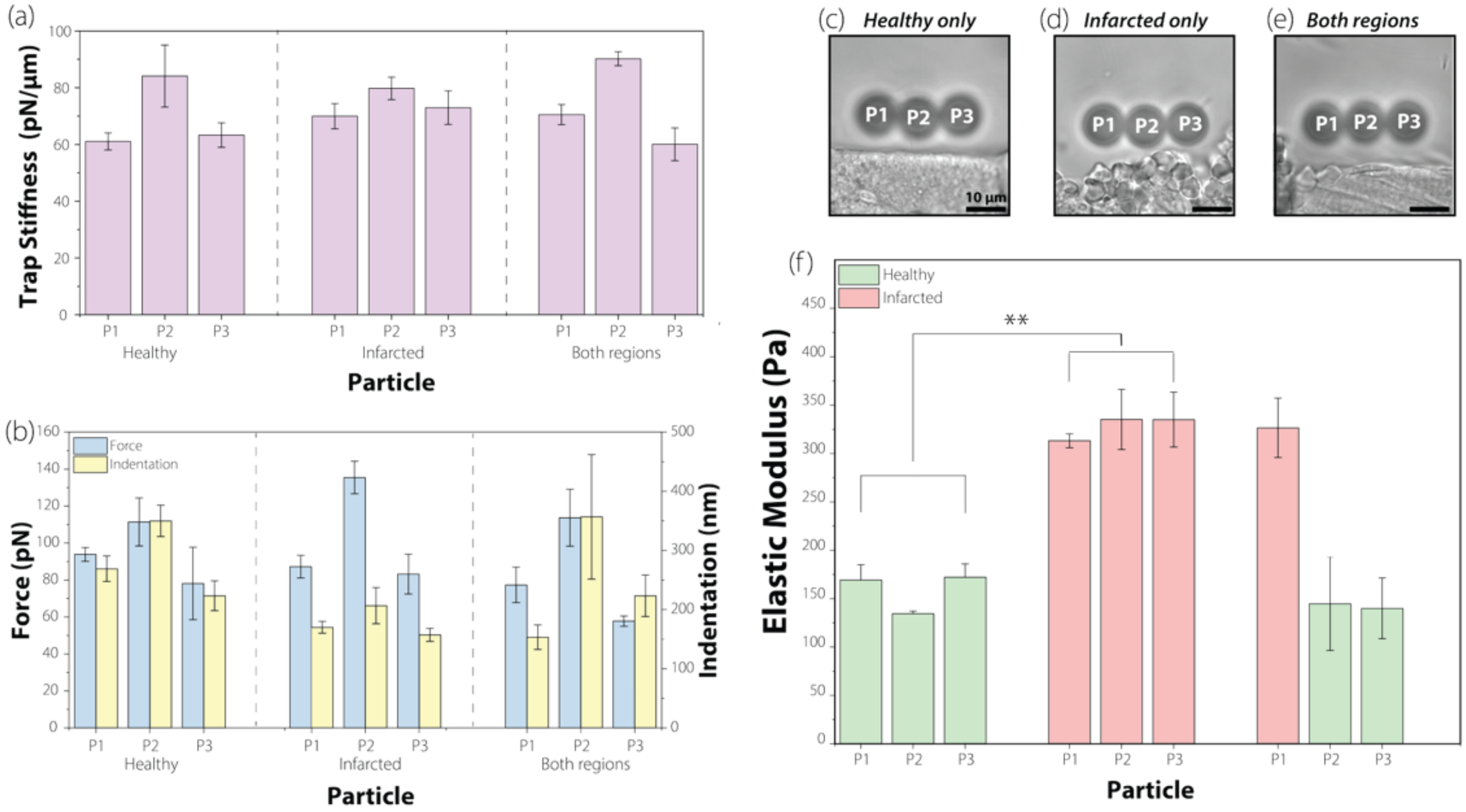

**Fig. 6.** LOFT indentation analyses on heterogenous myocardium samples. (a) Experimentally measured $\kappa$ for particles P1-P3 confined within LOFT under healthy-only, infarcted-only, and heterogeneous tissue samples. (b) Parallelized $\delta$ measured and applied forces for each particle within each condition. (c-e) Representative BF images of tissue topography. (f) Determined $E$ from each particle on tissue types. Particles that indent healthy or infarcted regions are labeled as green and red, respectively. Note ** represents significant differences indicated by a Student t-test (n = 3) by $p < 0.01$.

When comparing the healthy-only and infarcted-only tissue regions, average $\kappa$ across P1–P3 are found to be 74.64 ± 9.7 pN/ µm and 76.08 ± 6.46 pN/ µm respectively. During indentation of healthy tissue regions, the particles exert an average optical force of 94.64 ± 16.64 pN, producing an average $\delta$ of 280.76 ± 64 nm. In contrast, indentation experiments performed on infarcted tissue regions result in an average applied force of 101.95 ± 29 pN but elicits a substantially smaller average $\delta$ of 178 ± 25.62 nm. In the third configuration, P1 indents the infarcted region, whereas P2 and P3 simultaneously indent adjacent healthy regions. The $\kappa$ values range from approximately 66–82 pN/ µm, demonstrating moderate spatial variation in optical confinement across LOFT.

Particle P1 applies an average force of 77.37 pN, resulting in a $\delta$ of 153.33 nm. In comparison, P2 and P3 deliver an average force of 85.71 pN, producing significantly larger average indentations of approximately 290 nm. These observations further support the presence of locally increased mechanical stiffness within infarcted tissue relative to healthy myocardium [Figs. 6(a,b)].

For the homogeneous tissue experiments [Figs. 6(c,d)], the average $E$ obtained across all particles and experimental replicates are 158.7 ± 20.9 Pa for healthy tissue (green) and 327.8 ± 12.67 Pa for infarcted tissue (red). A $p$-value of 0.0096 was obtained demonstrating that LOFT can reliably differentiate tissue mechanical properties at low-force indentation regimes with statistical significance. In the heterogeneous tissue experiments [Fig. 6(e)], $E$ values obtained for P1, P2, and P3 are 326.55 ± 30.7 Pa, 144.78 ± 48.2 Pa, and 140.3 ± 31.3 Pa, respectively. Since P1 indents the infarcted region while P2 and P3 indent healthy regions, these measurements again demonstrated clear biomechanical separation between the two tissue types within a single experiment [Fig. 6(f)].

## 4 Discussion

Existing biomechanical characterization methods remain constrained by tradeoffs between force sensitivity, throughput, spatial resolution, and compatibility with dynamic *in situ* imaging. Although conventional OT uniquely operates within physiologically relevant pN force regimes, enabling minimally invasive mechanical interrogation with minimal sample perturbation, their broader application to tissue biomechanics has been limited by single-particle trapping and sequential force measurements captured by QPDs. These limitations are particularly significant for biological tissues, where mechanical properties can vary substantially over micron-scale distances due to local differences in cellular organization and ECM composition, and meaningful characterization frequently requires indentation depths extending beyond the displacement range for which conventional QPD-based measurements are optimized. Consequently, comprehensive characterization requires rapid, spatially resolved measurements that preserve the native mechanical state of the specimen. By integrating FLASH-UP, light-sheet illumination, and videography-based force transduction, LOFT enables simultaneous biomechanical characterization of multiple spatial locations while maintaining biologically relevant loading conditions. Beyond increasing throughput, this approach facilitates direct comparison of neighboring tissue regions under identical experimental conditions, minimizes measurement-induced mechanical perturbation, and supports highly reproducible and longitudinal studies.

Importantly, the current implementation represents only an initial demonstration of the platform's multiplexing capability, and its throughput can be further expanded through straightforward optical modifications. While the present setup increases biomechanical characterization throughput by threefold through simultaneous interrogation of three spatial locations, the throughput of LOFT is fundamentally governed by the spatial extent of the LS (Fig. S3) and probe size. Increasing the LS length would enable confinement of a greater number of dielectric probes, thereby proportionally increasing the number of simultaneous measurement sites. In the current optical configuration, the LS dimensions can be tailored by modifying the focal length of the cylindrical lens or by employing objective lenses with different NA and magnifications, both of which alter the extent of the illumination along the trapping axis or by using smaller-sized dielectric particles. These optical design parameters provide a straightforward route for scaling the platform toward higher-throughput, multipoint biomechanical characterization.

The utility of this approach was first demonstrated through single-point biomechanical characterization of both cellular and tissue systems. Macrophages revealed a substantially lower modulus compared to fibroblasts, consistent with the highly dynamic physiological role of macrophages, which require extensive migration through complex tissue environments during immune surveillance, inflammation, and wound healing. Such functionality necessitates a relatively compliant and deformable cellular phenotype capable of rapid shape adaptation and motility.[61,62] In contrast, fibroblasts are structural stromal cells responsible for ECM synthesis, collagen deposition, and maintenance of tissue architecture, resulting in stronger cytoskeletal organization and increased mechanical rigidity.[63,64] Furthermore, breast cancer cells demonstrated reduced stiffness and increased deformability which are commonly associated with malignant transformation because these properties facilitate migration, invasion, and metastatic progression through surrounding tissue microenvironments.[65] Healthy epithelial cells, in contrast, maintain stronger cell-cell and cell-matrix adhesions that contribute to a more mechanically stable phenotype.[66] Infarcted myocardium exhibited a higher $E$ than healthy tissue in accordance with previously reported literature demonstrating increased stiffness in infarcted myocardium due to fibrosis and pathological ECM remodeling.[9,10]

Having established agreement with well-characterized biomechanical differences across diverse biological models, we next evaluated whether LOFT could resolve spatially heterogeneous mechanical behavior within intact tissue specimens. The application of LOFT to myocardial tissue revealed statistically significant biomechanical differences between healthy and infarcted regions while simultaneously resolving localized heterogeneity within structurally complex specimens. These findings highlight an important advantage of parallelized optical force transduction: rather than treating tissues as mechanically homogeneous materials, LOFT enables interrogation of spatially varying biomechanical behavior within a single experiment. Since mechanical remodeling is intrinsically linked to alterations in tissue architecture, the integration of biomechanical measurements with quantitative 2PEF imaging and Fourier-domain microstructural analysis further establishes a framework for directly relating tissue mechanics to underlying structural organization. Such multimodal measurements are particularly relevant for diseases characterized by dynamic ECM remodeling, including fibrosis, cancer progression, and cardiovascular disease. Collectively, these findings demonstrate both the accuracy of LOFT for quantitative biomechanical measurements, and its ability to uncover spatially resolved structure-function relationships that are inaccessible using conventional OT.

## 4 Conclusion

In conclusion, this work establishes LOFT as a versatile, multifunctional ultrafast optical biomechanics platform that extends the capabilities of OT beyond traditional single-particle force measurements toward high-throughput, spatially resolved biomechanical interrogation of complex biological systems. By integrating femtosecond-enhanced trapping, light-sheet-mediated multiparticle confinement, videography-based force transduction, and multimodal structural imaging, LOFT enables simultaneous mechanical characterization and visualization under biologically compatible loading conditions while preserving native tissue architecture. To our knowledge, this work represents the first demonstration of OT-based mechanical testing performed directly on intact soft tissue specimens and the first implementation of parallelized optical force transduction for multipoint tissue biomechanics. Beyond improving experimental throughput, LOFT enables direct investigation of how local mechanical properties emerge from underlying tissue microstructure and how these structure-function relationships evolve during development, disease progression, remodeling, and regeneration. Looking forward, incorporation of a spatial light modulator would enable dynamic engineering of the LS intensity distribution and adaptive control of the optical force landscape. Unlike the static trapping geometry produced by a cylindrical lens, programmable wavefront shaping could tailor the trapping profile to complex tissue geometries and heterogeneous surface topographies, improving force uniformity across curved or irregular biological specimens while further increasing multiplexing capability. More broadly, the modular architecture of LOFT provides a foundation for future advances in adaptive optical manipulation and multimodal biomechanical imaging. As such, LOFT establishes a new direction for optical biomechanics by enabling minimally invasive, high-content characterization of spatially resolved structure-function relationships in living biological systems.

*Disclosures*

The authors declare that there are no financial interests, commercial affiliations, or other potential conflicts of interest that could have influenced the objectivity of this research or the writing of this paper.

*Code, Data, and Materials Availability*

All data needed to evaluate the conclusions in the paper are present in the paper and/or the Supplementary Material. Additional data related to this paper may be requested from the authors.

*Acknowledgments*

We thank Brown University for funding. We thank Wenyu Liu and Shayaan Chaudhary on helpful discussions on image processing

**Krishangi Krishna** received her Ph.D. and B.E. in Biomedical Engineering from Brown University and Stony Brook University, respectively. Her research interests lie in optical tweezers, ultrafast laser optics, and biophotonics.

**Kimani C. Toussaint, Jr.** is the Thomas J. Watson, Sr. Professor in the School of Engineering at Brown University. He is the senior associate dean in the School of Engineering and director of the Brown–Lifespan Center for Digital Health. He directs the laboratory for Photonics Research of Bio/Nano Environments, an interdisciplinary research group focusing on developing nonlinear optical imaging techniques for the quantitative assessment of biological tissues and novel methods for harnessing plasmonic nanostructures for light-driven control of matter.

## Caption List

**Fig. 1** Experimental arrangement. (a) Schematic illustration of LOFT. L; lens, SF; spatial filter, HWP; half wave plate, LP; linear polarizer, ND; neutral density, QWP; quarter wave plate, M; mirror, FM; flip mirror, SL; scan lens, TL; tube lens, CL; cylindrical lens, DM; dichroic mirror, SPF; short pass filter. (b) Experimentally obtained beam profile of Gaussian. (c) Normalized transverse intensity profile along x (blue) and y (red). (d) Transverse beam profile of LS. (e) Corresponding normalized intensity profiles along x (blue) and y (red).

**Fig. 2** Biomechanical characterization pipeline. (a) Representative BF time lapse demonstrating optical indentation of a macrophage cell (red) using a trapped dielectric microsphere (green). (b) Corresponding stage displacement (red), bead displacement (green), calculated indentation depth (blue), and applied force (orange). (c) Force-indentation curve illustrating approach (purple) and retraction (pink).

**Fig. 3** Single indentation results of cells. (a) Comparison of δ under comparable force applied for macrophages and fibroblasts. (b) Extrapolated E for macrophages and fibroblasts. (c) Comparison of δ measured for healthy breast epithelial and cancer cells. (d) Calculated E for breast epithelial and cancer cells. Note ** represents significant differences indicated by a Student's t-test ($n = 3$) by $p < 0.01$.

**Fig. 4** Single indentation results of myocardium. (a) Comparison of δ measured for both tissue types using 2D Gaussian OT. (b) Estimated E for healthy and infarcted tissue. Note * represents significant differences indicated by a Student's t-test ($n = 3$) by $p < 0.05$.

**Fig. 5** Multiphoton imaging and analysis to determine tissue microarchitecture. (a,b) BF visualization of healthy and infarcted tissues. (c,d) 2PEF images from the insets of (a,b). (e,f) Spatial map illustrating anisotropic, isotropic and dark regions identified from analysis. (g,h) Polar histograms illustrating fiber orientation distribution and CV measurements.

**Fig. 6** LOFT indentation analyses on heterogenous myocardium samples. (a) Experimentally measured κ for particles P1-P3 confined within LOFT under healthy-only, infarcted-only, and heterogeneous tissue samples. (b) Parallelized δ measured and applied forces for each particle within each condition. (c-e) Representative BF images of tissue topography. (f) Determined E from each particle on tissue types. Particles that indent healthy or infarcted regions are labeled as green and red, respectively. Note ** represents significant differences indicated by a Student t-test ($n = 3$) by $p < 0.01$.

**Fig. S1** Characterization of multiparticle trapping in LOFT. Positional probability distributions of three simultaneously trapped particles (P1–P3) measured along the (a) x- and (b) y-axes. The Gaussian distributions indicate stable optical confinement within LOFT.

**Fig. S2** Time-lapse sequence illustrating a representative LOFT lateral indentation cycle. The trapped bead array is translated toward the tissue (Approach), establishes contact with the tissue surface (Contact), and subsequently indents the tissue during continued stage displacement (Indentation).

**Fig. S3** Assuming a constant optical intensity, the required laser power scales linearly with the length of the light sheet, allowing a corresponding increase in the number of simultaneously trapped particles. The labels denote the predicted number of trapped particles for each light-sheet extent, where the star represents the current work's throughput.

**Video 1** Single indentation of macrophage cell (MPEG, 5 MB).
**Video 2** Single indentation of fibroblast cell (MPEG, 4.2 MB).
**Video 3** Single indentation of breast cancer cell (MPEG, 5 MB).
**Video 4** Single indentation of breast epithelial cell (MPEG, 4.2MB).
**Video 5** Single indentation of healthy myocardium (MPEG, 14.1 MB).
**Video 6** Single indentation of infarcted myocardium (MPEG, 13.7 MB).
**Video 7** Photonic jets during multiparticle trapping (MPEG, 10.1 MB).
**Video 8** LOFT multi-indentation of healthy myocardium (MPEG, 9.4 MB).
**Video 9** LOFT multi-indentation of infarcted myocardium (MPEG, 14.5 MB).
**Video 10** LOFT multi-indentation of heterogenous myocardium (MPEG, 45.3 MB).